\documentclass[10pt,twocolumn,letter]{IEEEtran}

\usepackage{amsmath,graphicx}
\usepackage{amsfonts}
\usepackage{amssymb}
\usepackage{epsfig}
\usepackage{mathrsfs}
\usepackage{graphicx}
\usepackage{algorithm}
\usepackage{caption}
\usepackage{subcaption}
\usepackage{epstopdf}
\usepackage{color}
\usepackage{cite}
\usepackage{balance}
\usepackage{subcaption}
\usepackage{mathtools}

\newcommand{\F}{{\cal F}}

\newcommand{\beq}{\begin{equation}}
\newcommand{\eeq}{\end{equation}}

\def\ba{\mbox{\boldmath $a$}}

\def\bee{\mbox{\boldmath $e$}}
 \def\bm{\mbox{\boldmath $m$}}

\def\bs{\mbox{\boldmath $s$}}

\def\bu{\mbox{\boldmath $u$}}
\def\bw{\mbox{\boldmath $w$}}

\def\bx{\mbox{\boldmath $x$}}

\def\br{\mbox{\boldmath $r$}}
\def\bs{\mbox{\boldmath $s$}}

\def\mA{\mbox{$\mathbf{A}$}}

\def\mC{\mbox{$\mathbf{C}$}}

\def\mI{\mbox{$\mathbf{I}$}}

\def\mU{\mbox{$\mathbf{U}$}}

\floatstyle{ruled}
\newfloat{algorithm}{tbp}{loa}
\providecommand{\algorithmname}{Algorithm}
\floatname{algorithm}{\protect\algorithmname}

\newtheorem{theorem}{Theorem}

\newtheorem{definition}{Definition}

\newenvironment{proof}[1][Proof]{\noindent \textbf{#1.} }{\qedsymbol}
\newcommand{\qedsymbol}{\hspace{\fill}\rule{1.5ex}{1.5ex}}

\makeatother

\begin{document}
	
	\title{Dynamic Resource Optimization for Decentralized Estimation in Energy Harvesting IoT Networks\vspace{.2cm}}
	
	\author{Claudio Battiloro,~\IEEEmembership{Student Member,~IEEE}, Paolo~Di Lorenzo,~\IEEEmembership{Senior Member,~IEEE},\medskip\\Paolo Banelli,~\IEEEmembership{Member,~IEEE}, and Sergio Barbarossa~\IEEEmembership{Fellow,~IEEE} \vspace{-.5cm}\thanks{Battiloro, Di Lorenzo and Barbarossa are with the Department of Information Engineering, Electronics, and Telecommunications, Sapienza University of Rome, Via Eudossiana 18, 00184, Rome, Italy. Banelli is with the Department of Engineering, University of Perugia, via G. Duranti 93, 06128, Perugia, Italy.	E-mail: {\tt battiloro.claudio@gmail.com, \{paolo.dilorenzo,sergio.barbarossa\}@uniroma1.it, paolo.banelli@unipg.it.} This work was supported by the H2020 EU/Taiwan Project 5G CONNI, Nr.~861459, by MIUR under the PRIN Liquid Edge contract, and by Sapienza University ``Bandi di Ateneo per la ricerca 2019.''	Part of this work was presented in the conference paper \cite{di2019dynamic}.}}

	\maketitle
	
\begin{abstract}
We study decentralized estimation of time-varying signals at a fusion center, when energy harvesting sensors transmit sampled data over rate-constrained links. We propose dynamic strategies to select radio parameters, sampling set, and harvested energy at each node, with the aim of estimating a time-varying signal while ensuring: i) accuracy of the recovery procedure, and ii) stability of the batteries around a prescribed operating level. The approach is based on stochastic optimization tools, which enable adaptive optimization without the need of apriori knowledge of the statistics of radio channels and energy arrivals processes. Numerical results validate the proposed approach for decentralized signal estimation under communication and energy constraints typical of Internet of Things (IoT) scenarios.
\end{abstract}

\smallskip
\begin{IEEEkeywords}
	Wireless sensor networks, decentralized estimation, quantization, energy harvesting, stochastic optimization.
\end{IEEEkeywords}

\section{Introduction}
Wireless sensor networks (WSNs) are envisioned to play a key role in the IoT paradigm, where a huge number of smart devices are expected to be connected with each other while sensing information from the environment \cite{stankovic2014research,Barb-Sard-Dilo}. In fact, thanks to  machine learning and signal processing tools, IoT networks will enable distributed proactive sensing and control mechanisms aimed at preventing performance degradation and optimizing the overall system. A key technical enabler to realize such vision is decentralized signal estimation, which was deeply investigated in several recent works as, e.g., \cite{Barb-Sard-Dilo,xiao2005decentralized,xiao2006power,li2007rate,barbarossa2013optimal,di2018optimal}. Decentralized estimation aims at gathering data collected by a WSN in a central node, i.e., the fusion center (FC), with the goal of evaluating globally optimal estimates of a signal of interest. Assuming the presence of realistic rate-constrained radio channels, data must necessarily be quantized before being transmitted to the FC, thus introducing an inevitable distorsion that reduces the performance of the signal estimation task. In this context, considering static parameter estimation, the works in \cite{xiao2005decentralized,xiao2006power,li2007rate} proposed to optimize the quantization scheme and the radio resource allocation while imposing a constraint on the mean-square error (MSE) performance. This approach was then extended also to decentralized detection in \cite{barbarossa2013optimal}, and to graph signal interpolation in \cite{di2018optimal}.

The aforementioned methods are inherently static, i.e., they do not consider possible temporal variation of the environment (e.g., the radio channels) or the signal of interest. Furthermore, energy constraints at each sensor were considered only over a single time snapshot, without keeping into account the overall lifetime of the WSN, which depends on the evolution of the batteries owned by each device. In the last years, energy harvesting (EH) techniques have attracted a lot of interest in IoT in order to cope with the battery-limited nature of sensor devices, thus enabling the possibility to collect energy from renewable sources such as wind, sun, vibration, and heat \cite{kuEH2016,priya2009energy}. EH naturally introduces dinamicity in the estimation problem due to the intermittent arrivals of energy from the environment and the variability over time of the battery levels.

In this context, a common approach to optimize performance is maximizing the throughput of the energy harvesting sensor communication system \cite{ulukusEHRew2016, maMQAM_EH_2015}. The work in \cite{blascoMDP_EH_2013} exploited a Markov decision process (MDP) procedure to maximize the long-term expected throughput and get the optimal power level. However, the large cardinality of the state and action space makes the computational complexity of the MDP-based approaches generally high. An energy scheduling strategy for remote estimation in the case of a single EH sensor was also proposed in \cite{nayyar2013optimal}. The work in \cite{rezaeeEH_Contin_ene} considered the single-user throughput
maximization of an energy harvesting system with continuous energy and data arrivals. The works in \cite{ku2015data,sharma2010optimal} studied the optimal packet communication strategy to maximize the net bit rates while stabilizing the data queue in EH communications. With the same aim, a packet communication strategy to maximize the net bit rates in EH communications is proposed also in \cite{qiuLyap2018}, but constrained on bounded long-term average battery level and bit error rate. The work in \cite{marano2018} proposed an energy scheduling strategy to maximize the total information collected by an agent in the SENMA (Sensor Networks with Mobile Agents) paradigm. Finally, the work in \cite{zhou2016optimal} proposed a dynamic radio resource allocation for static and dynamic estimation in WSNs with EH devices, in the case of scalar parameter estimation and analog amplify-and forward transmission strategies.

\textbf{Contributions.} In this paper, we study decentralized estimation in EH WSNs, proposing optimal dynamic resource allocation strategies to strike the best possible tradeoff between accuracy of the estimation task and energy spent by the WSN. Differently from the previous works \cite{ulukusEHRew2016, maMQAM_EH_2015,blascoMDP_EH_2013,rezaeeEH_Contin_ene,ku2015data,sharma2010optimal,qiuLyap2018}, we do not consider throughput as our main optimization objective. Instead, similarly to \cite{zhou2016optimal}, we study the energy-accuracy tradeoff that we have in decentralized signal estimation, taking explicitly into account estimation performance (e.g., mean-square error) as our main objective (or constraint). However, differently from \cite{zhou2016optimal}, we consider a \textit{vector} parameter estimation problem, and we assume the presence of a \textit{digital quantizer} that performs analog to digital conversion at each device. In particular, the contribution of this paper is twofold:
\begin{enumerate}
    \item We devise a dynamic strategy that optimally selects the radio parameters (i.e., transmission energies, bits), the set of sampling sensors, and the amount of harvested energies in order to maximize the signal estimation performance, while imposing specific guarantees in terms of stability of the battery levels.\vspace{.1cm}
    \item We propose a dynamic algorithm that minimizes the average energy expenditure of the WSN, while imposing prescribed performance guarantees in terms of average estimation performance and stability of the batteries.
\end{enumerate}
The proposed techniques guarantee that the sensor batteries always lie into an operating region that prevents the network from running out of energy. As we will see in the sequel, in the first case, this property is guaranteed by accepting a proper tradeoff between the performance of the signal estimation task and the size of the batteries at each sensor. On the other hand, the second technique aims at striking an optimal tradeoff between network energy expenditure (and, as a byproduct, number of transmitting nodes) and average performance of the signal recovery task. Our dynamic optimization strategies automatically selects also the set of sampling (and transmitting) nodes, i.e., those devices that collect data and transmit them using at least one bit of information. The proposed strategies hinge on stochastic and Lyapunov optimization techniques \cite{neely2010stochastic}, which enable to learn the best resource allocation over time by simply observing instantaneous realizations of the energy arrivals and the radio channels, without any previous knowledge of the statistics of these random processes. Numerical results illustrate the validity of the proposed approach by assessing its performance in several practical scenarios.


\textbf{Outline.} The paper is organized as follows. In Sec. II, we introduce the system model, comprising the adopted probabilistic quantization scheme, Bayesian estimator, and energy harvesting model. Then, in Sec. III, we develop an algorithmic solution that dynamically optimizes estimation accuracy while guaranteeing stability of the devices batteries. In Sec. IV, we develop a dynamic algorithm aimed at minimizing the network energy expenditure under estimation accuracy and battery stability constraints. Finally, Sec. V draws some conclusions.

\textbf{Notation.} Scalar, column vector, and matrix variables are respectively indicated by plain letters $a$ ($A$), bold lowercase letters $\ba$, and bold uppercase letters $\mA$. $\mathbb{I}(\cdot)$ denotes the indicator function; $a_{ij}$ is the $(i, j)$-th element of $\mA$, $\mI$ is the identity matrix, and $\mathbf{1}_N$ ($\mathbf{0}_N$) is the $N\times 1$ vector of all ones (zeros).  ${\rm diag}\{\ba\}$ denotes a diagonal matrix having vector $\ba$ on its main diagonal. $\mathbb{E}\{\cdot\}$ denotes the expectation operator. $\mathrm{Tr}\{\cdot\}$ denotes the matrix trace operator. Other specific notation is defined along the paper, whenever it is needed.

\section{System Model}
Let us consider a WSN with $N$ nodes that is deployed to monitor a signal of interest over a certain geographic area. We consider a dynamic scenario where time is divided in slots of equal duration $T$. Let $\bx(t) = [x_1(t),...,x_N(t)]^T$ be the vector collecting the signal values measured by all the nodes of the network at time $t$. The gathered measurements may be highly unreliable due to observation noise, presence of outliers, missing data, etc. Improving the reliability of the individual node is typically unfeasible because of increased complexity and cost, which are fundamental design constraints in large scale networks. A way to recover reliability is to properly fuse the measurements collected over all the network in order to reach some globally optimal decision. This is possible if the set of data gathered  by the network exhibits some kind of {\it structure} (e.g., correlations, dependencies, smoothness, etc.), which is typically the case in many physical fields of interest, e.g., the distribution of temperatures or the concentration of a contaminant. In mathematical terms, this means that the observed signal field belongs to a low-dimensional subspace, i.e., the vector $\bx(t)$ can be modeled as:
\begin{gather}\label{vector_obs}
\bx(t) = \mathbf{U}\bs(t),
\end{gather}
where $\mathbf{U}$ is an $N \times r$ matrix, with $r \leq N$, and $\bs(t)$ is an $r \times 1$ column vector. The columns of $\mU$ are assumed to be linearly independent and thus constitute a basis spanning the signal subspace. In many applications, the signal is a smooth function, which can be very well modeled by choosing the columns of $\mU$ as the low frequency components of the Fourier basis, or low-order polynomials, for example. In practice, the dimension $r$ of the signal subspace is typically much smaller than the dimension $N$ of the observation space \cite{vetterli2014foundations,bishop2006pattern}.
\subsection{Probabilistic Quantization}
From (\ref{vector_obs}), at time $t$, the network collects noisy measurements $\{y_i(t)\}_{i=1}^{N}$ given by:
\begin{gather}
\label{observation}
y_i(t) = x_i(t) + v_i(t) = \bu_i^T\bs(t) + v_i(t),
\end{gather}
$i=1,...,N$, where $\bu_i^T$ is the $i$-th row of the matrix $\mathbf{U}$, and $v_i(t)$ is zero-mean, uncorrelated noise with variance $\sigma_i^2$. The measurements in (\ref{observation}) must be transmitted to an FC to evaluate an optimal estimate for the signal $\bx(t)$. Then, assuming the presence of rate-constrained radio channels, the messages $\{y_i(t)\}$ must necessarily be encoded into a sequence of bits in order to be sent to the FC. Suppose that $[-A,A]$ is the signal range that sensors can observe. At each time $t$, we consider a uniform quantizer at each node $i$, which divides the range $[-A,A]$ into
intervals of length $\Delta_i(t)=2A/(2^{b_i(t)}-1)$, and rounds the observations in \eqref{observation} to the neighboring endpoints of these intervals in a probabilistic manner \cite{xiao2005decentralized}, \cite{xiao2006power}. Then, if $$l\Delta_i(t) \leq y_i(t) \leq (l+1)\Delta_i(t),$$ with $l\in\{-2^{b_i(t)-1},...,2^{b_i(t)-1}\}$, $y_i(t)$ is quantized to $m(y_i(t),b_i(t))$ according to:
\begin{gather}\label{quant}
m(y_i(t),b) = l\Delta_i(t)+\alpha\Delta_i(t),
\end{gather}
where $\alpha$ is a Bernoulli random variable such that $\mathbb{E}\{\alpha\}={\rm Prob}\{\alpha=1\}=(y_i(t)-l\Delta_i(t))/\Delta_i(t)\in [0,1]$. Thus, according to (\ref{quant}) and (\ref{observation}), the $i$-th quantized  observation at time $t$, i.e., $m_i(y_i(t),b_i(t))$, can be equivalently written as:
\begin{gather}\label{quant_observation}
m_i(y_i(t),b_i(t))= \bu_i^T\bs(t) + v_i(t)+q(y_i(t),b_i(t)),
\end{gather}
where $q(y_i(t),b_i(t))=(\alpha-\mathbb{E}\{\alpha\})\Delta_i(t)$ denotes the zero-mean quantization noise with variance
\begin{equation} \label{quant_variance}
  \sigma^2_q(b_i(t))=\displaystyle \frac{A^2}{(2^{b_i(t)}-1)^2},\quad i=1,\ldots,N.
\end{equation}

\subsection{Bayesian LMMSE Estimation over Rate-constrained Links}

To compute an optimal estimate of the signal, we consider a Bayesian approach, where the vector $\bs(t)$ in (\ref{quant_observation}) is assumed to be a random process with mean $\boldsymbol{\mu}_s$ and covariance matrix $\mC_{s}$ \cite{kay1993fundamentals}. Also, from (\ref{quant_observation}), the additive disturbance comprises both observation and quantization noises, i.e.,
$$w_i(t) = v_i(t)+q(y_i(t),b_i(t)),$$ $i=1,\ldots,N$. Thus, letting $\bw(t) = [w_1(t),...,w_N(t)]^T$, since observation and quantization noise are uncorrelated, the total disturbance $\bw(t)$ is a zero-mean random vector with covariance matrix [cf. (\ref{quant_variance})]:
\begin{align}\label{noise_covariance}
    \mC_{w}
    &={\rm diag}\left\{\left\{\sigma_i^2+\frac{A^2}{(2^{b_i(t)}-1)^2}\right\}_{i=1}^N\right\}.
\end{align}
Furthermore, the noise term $\bw(t)$ is assumed to be uncorrelated from the signal process $\bs(t)$. Under such assumptions, the Bayesian Gauss-Markov Theorem defines the best linear minimum mean-square error (LMMSE) estimator (if $\bs(t)$ is a Gaussian process, it is also optimal among all estimators) for $\bs(t)$, which reads as \cite[p.391]{kay1993fundamentals}:
\begin{equation}\label{lmmse}
\widehat{\bs}(t)=\boldsymbol{\mu}_s+\left(\mC_{s}^{-1}+\mU^T\mC_w^{-1}\mU\right)^{-1}\mU^T\mC_w^{-1}(\bm(t)-\mU\boldsymbol{\mu}_s)
\end{equation}
where $\bm(t) = [m(y_1(t),b_1(t)),...,m(y_N(t),b_N(t))]^T$. The performance of the LMMSE estimator is measured by the error $\boldsymbol{\varepsilon}(t)=\widehat{\bs}(t)-\bs(t)$, which is zero-mean and has covariance matrix given by
$\mC_{\varepsilon}=\left(\mC_s^{-1}+\mU^T\mC_w^{-1}\mU\right)^{-1}$. In particular, the Bayesian Mean-Square Error (BMSE) achieved by the LMMSE estimator in (\ref{lmmse}) is given by:
\begin{align}\label{BMSE}
{\rm BMSE} &= \mathbb{E}\|\boldsymbol{\varepsilon}(t)\|^2={\rm Tr} \left\{ \left(\mC_s^{-1}+\mU^T\mC_w^{-1}\mU\right)^{-1}\right\}\nonumber\\
&\hspace{-.4cm}= {\rm Tr} \left\{ \left(\mC_s^{-1}+   \sum_{i=1}^N \,\frac{\boldsymbol{u}_{i}\boldsymbol{u}_{i}^T}{\displaystyle\sigma_i^2+\frac{A^2}{(2^{b_i(t)}-1)^2}}   \right)^{-1}\right\},
\end{align}
where the last equality in (\ref{BMSE}) follows from (\ref{noise_covariance}).

The BMSE expression in (\ref{BMSE}) depends on the bits $\{b_i(t)\}_{i=1}^N$ used by the nodes to quantize the measured signal. We will now relate the BMSE expression in (\ref{BMSE}) to the energy needed to transmit such amount of information within each time slot of duration $T$. To this aim, let us assume that the channel between each sensor and the FC is corrupted with additive white Gaussian noise, whose double-sided power spectrum density is given by $N_0/2$. Furthermore, we denote by $h_i(t)$ the channel coefficient between sensor $i$ and the FC at time $t$. If sensor $i$ sends $b_i(t)$ bits in a time slot of duration $T$, using quadrature amplitude modulation with constellation size $2^{b_i(t)}$ at a bit error probability ${\rm BER}_i$, then the amount of energy required for the transmission is \cite{xiao2006power,cui2005energy,cui2004joint}:
\begin{align}
\label{power_QAM}
e_i(t)=\frac{2N_f N_0G_d T}{h_i^2(t)}\left( \ln\frac{2}{\rm BER_i}\right)(2^{b_i(t)}-1),
\end{align}
where $N_f$ is the receiver noise figure, and $G_d$ is a system constant defined in the same way as in \cite{cui2005energy,cui2004joint}. In the sequel, for simplicity, we assume that the BER of each transmission is made sufficiently small such that transmission errors have a negligible effect on the BMSE. Thus, letting $$c_i(t)=\frac{2N_fN_0G_d T}{h_i^2(t)}\left( \ln\frac{2}{\rm BER_i}\right),$$ and using (\ref{power_QAM}) in (\ref{BMSE}), the BMSE is given by:
\begin{align}
\label{BMSE2}
\hspace{-.15cm}{\rm BMSE}(\bee(t))={\rm Tr}\left\{ \left(\mathbf{C}_{s}^{-1} + \sum_{i=1}^N \,\frac{\bu_{i}\bu_{i}^T}{\displaystyle\sigma_i^2+\frac{A^2c_i^2(t)}{e_i^2(t)}}\right)^{-1} \right\}
\end{align}
where $\bee(t)=[e_1(t),\ldots,e_N(t)]^T$ is the vector collecting all transmission energies. If no node is transmitting, signal estimation is still performed as in \eqref{lmmse}, but exploiting only the prior information on the signal process. In such a case, we have $\bm(t)=\mathbf{0}$, $\mC_w=\boldsymbol{\Sigma}={\rm diag}\{\sigma_1^2,\ldots,\sigma_N^2\}$, and the LMMSE estimator reads as:
\begin{equation}\label{lmmse_no_nodes}
\widehat{\bs}(t)=\left(\mI - \left(\mC_{s}^{-1}+\mU^T\boldsymbol{\Sigma}^{-1}\mU\right)^{-1}\mU^T\boldsymbol{\Sigma}^{-1}\mU\right)\boldsymbol{\mu}_s,
\end{equation}
with corresponding BMSE (i.e., the worst case) given by:
\begin{align}
\label{BMSE_no_nodes}
\hspace{-.15cm}{\rm BMSE}^{worse}={\rm Tr}\left\{\mathbf{C}_{s} \right\}.
\end{align}

\subsection{Energy Harvesting Model}

The EH process is modeled as successive energy packet arrivals, i.e., $R_i(t)$ units of energy arrive at sensor $i$ at the beginning of the $t$-th time slot. The energy arrivals $R_i(t)$ are i.i.d. among different slots, and are upper bounded by $R^{max}$ \cite{huang2013utility}. In each time slot, part of the arrived energy, say, $r_i(t)$, satisfying $r_i(t) \leq R_i(t)$, will be harvested and stored in the battery, and it will be available for transmission [cf. (\ref{power_QAM})] from the next slot. Let us denote the battery level of node $i$ at time slot $t$ as $B_i(t)$. At each time $t$, the battery is drained by node's transmissions toward the fusion center, which incur an energy cost $e_i(t)$, and by other operations made by the node during the slot (such as, e.g., processing, signalling, etc.), which require an energy cost $e_{o,i}$. In the sequel, we assume that $e_{o,i}$ is fixed over time and known apriori (i.e., it depends on the structure of the specific algorithmic solution). Then, the system has an implicit energy causality constraint $e_i(t) \leq B_i(t) - e_{o,i}$ for all $t$, which let the battery evolve according to:
\begin{gather}\label{battery}
B_i(t+1)=B_i(t) -e_i(t) - e_{o,i}+ r_i(t),\; \text{ for all } i,t.
\end{gather}
The energy causality constraint ensures that $B_i(t) \ge 0$ for all $i, t$.  Also, in (\ref{battery}), we omit the maximum battery size because our methods will stabilize the batteries with guaranteed upper-bounds on their levels. Of course, from (\ref{battery}), the battery level is determined by the balance between the energy spent for transmission/processing [i.e., $e_i(t)$ and $e_{o,i}$] and the one harvested from the environment [i.e., $r_i(t)$]. In the sequel, we illustrate the proposed methods for decentralized estimation based on stochastic Lyapunov optimization.

\section{Dynamic BMSE Minimization under Battery Stability Constraints}

The proposed strategy aims at minimizing the temporal average of the BMSE in (\ref{BMSE2}), constrained to the aforementioned EH and battery processes, with respect to the transmission energies $\bee(t)=\{e_i(t)\}_{i=1}^N$ and the harvestable energies $\br(t)=\{r_i(t)\}_{i=1}^N$. In principle, since the energies $\{e_i(t)\}_{i=1}^N$ are related to the quantization bits $\{b_i(t)\}_{i=1}^N$ through (\ref{power_QAM}), the values of $\{e_i(t)\}_{i=1}^N$ belong to a finite discrete set. This would inevitably lead to a mixed-integer problem formulation, which has a prohibitive (combinatorial) complexity for large number of nodes and quantization levels. Thus, to find an approximate but low-complexity solution, we relax $\{e_i(t)\}_{i=1}^N$ to be real variables, and then we cast our optimization problem as:
\begin{align}\label{Problem}
&\operatorname*{min}_{\boldsymbol{e}(t),\,\boldsymbol{r}(t)}\;\; \lim_{t\to \infty} \frac{1}{t} \sum_{\tau=0}^{t-1} \mathbb{E}\{{\rm BMSE}(\bee(\tau))\}
\\& \quad\qquad \text{subject to} \nonumber\\
&\quad \qquad \qquad \lim_{t\to \infty} \frac{1}{t} \sum_{\tau=0}^{t-1} \mathbb{E}\{B_i(\bee(\tau))\}<\infty \quad \forall i;\nonumber\\
&\quad\qquad \qquad 0\leq e_i(t) \leq \text{min}[e_i^{max}, B_i(t)-e_{o,i}] \quad\forall i,t; \nonumber
\\& \quad \qquad \qquad 0\leq r_i(t) \leq R_i(t) \quad\forall i,t. \nonumber
\end{align}
The first constraint in (\ref{Problem}) imposes that the batteries are strongly stable \cite{neely2010stochastic}, i.e., they cannot grow unbounded; the second constraint puts bounds on the transmitted powers, whose maximum value is given by the minimum among the battery level $B_i(t)$ minus the overhead energy $e_{o,i}$ and the maximum energy $e_i^{max}$ that can be transmitted by the radio interface; finally, the third constraint sets the bounds on the harvestable energy at each time slot.

To solve problem (\ref{Problem}), we employ dynamic methods based on stochastic optimization \cite{neely2010stochastic,neely2010dynamic}. In particular, to guarantee the energy causality constraint $e_i(t) \leq B_i(t)-e_{o,i}$, for all $i$, $t$, and keep the energy storage (strongly) stable around a prescribed battery level, we use the approach from \cite{neely2010dynamic}, defining the virtual queues:
\begin{gather} \label{Bi_tilde}
\tilde{B}_i(t) = B_i(t) - \theta_i,\qquad i = 1,...,N,
\end{gather}
where $B_i(t)$ evolves as in (\ref{battery}), and $\theta_i>0$ is a parameter to be selected. As illustrated in \cite{neely2010dynamic,mao2016dynamic}, the use of the virtual queues $\tilde{B}_i(t)$ in (\ref{Bi_tilde}) is useful to stabilize the battery levels $B_i(t)$ in (\ref{battery}) around $\theta_i$. Then, the algorithmic approach passes through the definition of the Lyapunov function:
\begin{gather}\label{Lyapunov1}
L\big(\mathbf{\tilde{B}}(t)\big) = \frac{1}{2} \sum_{i=1}^N\tilde{B}_i(t)^2
\end{gather}
where $\mathbf{\tilde{B}}(t)=\left\{\tilde{B}_i(t)\right\}_{i=1}^N$, and the corresponding one-slot conditional Lyapunov drift, given by:
\begin{gather}\label{Lyapunov_drift1}
\Delta\big(\mathbf{\tilde{B}}(t)\big)\triangleq\, \mathbb{E}\left\{L\big(\mathbf{\tilde{B}}(t+1))-L(\mathbf{\tilde{B}}(t)\big)\;\big|\;\mathbf{\tilde{B}}(t)\right\},
\end{gather}
where the expectation depends on the control policy, and is taken with respect to the random radio channels and energy packet arrivals. Then, since the problem formulation in (\ref{Problem}) aims at minimizing the  average BMSE, we introduce the drift-plus-penalty function defined as \cite{neely2010stochastic}:
\begin{gather}\label{Lyapunov_drift_p1}
\Delta_p\big(\mathbf{\tilde{B}}(t)\big) = \Delta(\mathbf{\tilde{B}}(t)) + V\cdot \mathbb{E}\left\{ {\rm BMSE}\big(\mathbf{e}(t)\big)\;\big|\;\mathbf{\tilde{B}}(t)\right\}
\end{gather}
where $V$ is a control parameter used to tradeoff the BMSE with batteries' size. Following arguments similar to those used in \cite[Lemma 4.6]{neely2010stochastic}, exploiting (\ref{Bi_tilde}), simple algebra shows that the drift-plus-penalty in (\ref{Lyapunov_drift_p1}) can be upper-bounded as:
\begin{align}\label{Lyapunov_drift_p_bound1}
\Delta_p\big(\mathbf{\tilde{B}}(t)\big) &\leq\, C+\sum_{i=1}^N\mathbb{E}\left\{\tilde{B}_i(t) \big[r_i(t)-e_i(t) \big]\;\big|\;\mathbf{\tilde{B}}(t)\right\} \nonumber\\
& \quad+V\cdot\mathbb{E}\left\{{\rm BMSE}(\mathbf{e}(t))\;\big|\;\mathbf{\tilde{B}}(t)\right\}
\end{align}
where $C$ is a positive constant. Now, we proceed by using a stochastic approach, where we drop the expectation and greedily minimize instantaneous values of (\ref{Lyapunov_drift_p_bound1}) at each $t$. Thus, in each time slot, the method requires the solution of the following optimization problem:
\begin{align} \label{nonconv_prob}
		&\operatorname*{min}_{\mathbf{e}(t),\mathbf{r}(t)}\quad \sum_{i=1}^N\tilde{B}_i(t)\,\big[r_i(t)-e_i(t)\big]  +V\cdot{\rm BMSE}(\mathbf{e}(t))\nonumber\\& \quad\qquad \text{subject to}
\\&\quad\qquad \qquad 0\leq e_i(t) \leq \text{min}[e_i^{max}, B_i(t)-e_{o,i}] \quad\forall i,t; \nonumber
\\& \quad \qquad \qquad 0\leq r_i(t) \leq R_i(t) \quad\forall i,t. \nonumber
\end{align}
The dynamic optimization of (\ref{nonconv_prob}) faces two main issues. The first issue is the nonconvexity of Problem (\ref{nonconv_prob}), due to the fact that the BMSE is a nonconvex function of the transmission energies $\mathbf{e}(t)$. The second issue is the presence of the batteries $\{B_i(t)\}_{i=1}^N$ into the optimization set, which makes the set non i.i.d. over time slots (a fact that we would like to exploit to prove convergence of the proposed algorithmic framework based on stochastic optimization, see \cite{neely2010dynamic}).

To tackle the first issue, we proceed by finding a suitable approximation of the objective function in \eqref{nonconv_prob} in order to simplify the solution while still preserving optimality guarantees. In particular, we hinge on the concept of \textit{$\Gamma$-additive approximation} \cite[p. $59$]{neely2010stochastic}, whose  definition is reported next.

\vspace{.1cm}
\begin{definition}
For a given constant $\Gamma$, a $\Gamma$-\textit{additive approximation}  of the drift-plus-penalty algorithm is one that, for a given state  $\mathbf{\tilde{B}}(t)$ at slot $t$, chooses a (possibly randomized) action $[\mathbf{e}(t),\mathbf{r}(t)]\in \mathcal{Z}(t)$ [cf. (\ref{set_Z})] that yields a conditional expected value of the objective function in \eqref{nonconv_prob} that is within a constant $\Gamma$ from the infimum over all possible control actions.
\end{definition}
\vspace{.1cm}

To find a suitable $\Gamma$-additive approximation, while preserving meaningful first-order information of the original function, we propose to substitute ${\rm BMSE}(\mathbf{e}(t))$ in \eqref{nonconv_prob} with its linearization around $\mathbf{e}(t-1)$ (i.e., the solution available at time $t-1$), which reads as:
\begin{align}\label{Approx_penalty}
\widehat{{\rm BMSE}}(\mathbf{e}(t))&\;=\;{\rm BMSE}(\mathbf{e}(t-1))\;+\nonumber\\ &\nabla_{\mathbf{e}}{\rm BMSE}(\mathbf{e}(t-1))^T(\mathbf{e}(t)-\mathbf{e}(t-1)),
\end{align}
where $\displaystyle\nabla_{\mathbf{e}}{\rm BMSE}(\mathbf{e}(t-1))=\left\{\frac{\partial {\rm BMSE}(\mathbf{e}(t-1))}{\partial e_i}\right\}_{i=1}^N$ is the gradient vector of the BMSE evaluated in $\mathbf{e}(t-1)$.

To tackle the second issue in \eqref{nonconv_prob}, we simply drop the batteries $\{B_i(t)\}_{i=1}^N$ from the optimization set. At first sight, this might seem an excessive approximation that can lead to a violation of the energy causality constraint $e_i(t)\leq B_i(t)- e_{o,i}$, for all $i=1,\ldots,N$. However, we will show that, under a proper choice of the system parameters $\theta_i$ in (\ref{Bi_tilde}) for all $i$, the proposed method can satisfy the energy causality constraint without explicitly consider it into the optimization. Then, the new optimization set becomes:
\begin{align}\label{set_Z}
&\mathcal{Z}(t) = \Big\{\mathbf{e}(t),\mathbf{r}(t)\;:\; 0\leq e_i(t) \leq e_i^{max} \nonumber
\\& \qquad \qquad 0\leq r_i(t) \leq R_i(t), \quad \forall i=1,\ldots,N \Big\},
\end{align}
which is now i.i.d. over time slots. Thus, using (\ref{Approx_penalty}) and (\ref{set_Z}) in (\ref{nonconv_prob}), we obtain the dynamic resource allocation policy given by the online optimization of the following per-slot problem (where we have removed all terms that do not depend on the optimization variables):
\begin{align}\label{main_prob}
&\operatorname*{min}_{\mathbf{e}(t),\mathbf{r}(t)}\quad \sum_{i=1}^N \bigg[-\tilde{B}_i(t)+V\cdot \frac{\partial {\rm BMSE}(\mathbf{e}(t-1))}{\partial e_i}\bigg]  e_i(t)\nonumber\\ &\qquad \qquad \quad + \sum_{i=1}^N\tilde{B}_i(t)r_i(t)\\
& \quad\qquad \text{subject to} \nonumber
\\&\quad\qquad \qquad 0\leq e_i(t) \leq e_i^{max} \quad\forall i,t; \nonumber
\\& \quad \qquad \qquad 0\leq r_i(t) \leq R_i(t) \quad\forall i,t. \nonumber
\end{align}
Problem \eqref{main_prob} is linear, and its globally optimal solution can be easily found in closed form, determining for every $t$ the values of the
energies $r_i(t)$ to be harvested from the environment, the transmission energies $e_i(t)$, and the sampling set
$$\mathcal{S}(t)=\{i\;:\; e_i(t)>0\},$$ i.e., the set of transmitting nodes at time $t$. In particular, minimizing (\ref{main_prob}) with respect to $r_i(t)$, with the constraint $0 \leq r_i(t) \leq R_i(t)$, node $i$ collects the maximum harvestable energy $R_i(t)$ when $B_i(t) \leq \theta_i$;
whereas, for $B_i(t) > \theta_i$, node $i$ does not harvest any energy. In formuals, we have:
\begin{equation}
    r_i(t) = R_i(t) \cdot \mathbb{I}(B_i(t) \leq \theta_i) \quad\forall\; i,t, \label{ri}
\end{equation}
where $\mathbb{I}(\cdot)$ denotes the indicator function. Similarly, since (\ref{main_prob}) is
linear with respect to $e_i(t)$, each node $i$ transmits using the maximum energy $e_i^{max}$ when $$B_i(t) \geq \theta_i+V\cdot \frac{\partial}{\partial e_i}{\rm BMSE}(\mathbf{e}(t)),$$ i.e., when the battery level is sufficiently high; in the opposite case, node $i$ does not transmit. Overall, we have:
\begin{equation}
e_i(t) = e_i^{max} \cdot \mathbb{I}\left(\tilde{B}_i(t) \geq V\cdot \frac{\partial}{\partial e_i}{\rm BMSE}(\mathbf{e}(t-1))\right) \;\;\forall\; i,t. \label{ei}
\end{equation}
The control policy achieved by the proposed Min-Drift-Plus Penalty strategy, together with the overall decentralized estimation procedure in EH-WSNs, are listed in Algorithm 1. The proposed method comes with theoretical guarantees, which we summarize in the following theorem.

\begin{algorithm}[t]
\caption*{\textbf{Algorithm 1: Dynamic BMSE minimization under battery stability constraints}}
\vspace{.1cm}
\textbf{Data:} Set $V>0$; $B_i(0)\geq e_i^{max}+2e_{o,i}$, $e_i(-1)$ chosen at random in $[0,e_i^{max}]$, $\theta_i = VG^{max} + 2e_i^{max}+2e_{o,i}$, for all $i=1,\dots,N$. Then, for each time $t\geq0$, observe the random events $\{h_i(t)\}_{i=1}^N$ and $\{R_i(t)\}_{i=1}^N$, and repeat:\smallskip
\begin{itemize}
\item \textbf{Nodes:} Compute harvested energies $\{r_i(t)\}_{i=1}^N$ and transmitted energies $\{e_i(t)\}_{i=1}^N$ as (\ref{ri}) and (\ref{ei}), respectively. Nodes belonging to the sampling set $\mathcal{S}(t)$ sense data from the environment, and transmit them to the fusion center. All nodes update batteries $\{B_i(t)\}_{i=1}^N$ as (\ref{battery}).\smallskip

\item \textbf{Fusion center:} Collect data from the sampling set $\mathcal{S}(t)$, and compute the LMMSE estimator in (\ref{lmmse}), where $m(y_i(t),b_i(t))$ and the $i$-th diagonal element of $\mC_w$ in (\ref{noise_covariance}) are set to 0 for all $i\notin \mathcal{S}(t)$.  Evaluate and transmit $\displaystyle\frac{\partial}{\partial e_i}MSE(\mathbf{e}(t))$ to node $i$, for all $i=1,\ldots,N$.
\end{itemize}
\end{algorithm}

\begin{theorem}\label{theorem}
Suppose that random radio channels $\{h_i(t)\}_{i,t}$ and energy packet arrivals $\{R_i(t)\}_{i,t}$ are i.i.d over time, and that $\mathbb{E}\{L(\tilde{\mathbf{B}}(0))\}<\infty$. Also, let $B_i(0)\geq e_i^{max}+2e_{o,i}$, and
\begin{equation}\label{theta}
    \theta_i= VG_i^{max}+2e_i^{max}+2e_{o,i}, \quad i=1,\ldots,N,
\end{equation}
where
\begin{align}\label{Gimax}
    &G_i^{max}=  \frac{1}{2e_i^{max}\sigma_i^2} {\rm Tr}\left\{ \left(\mathbf{C}_{s}^{-1} + \sigma_i^{-2}\bu_i\bu_i^T\right)^{-2}\hspace{-.1cm}\bu_i\bu_i^T  \right\},
\end{align}
 for all $i=1,\ldots,N$. Then, the dynamic control policy obtained by Algorithm 1 guarantees the following properties.
\begin{description}
\item[(a)] \emph{\texttt{[Battery evolution]}}: The battery level at each node satisfy:
\begin{equation}\label{batt1}
e_i^{max}+e_{o,i}  \leq B_i(t) \leq \theta_i+R^{max}-e_{o,i}, \quad \forall i,t,
\end{equation}
i.e., the batteries are strongly stable and never drain. From (\ref{theta}) and (\ref{batt1}), it also holds:
\begin{equation}\label{batt2}
    B_i(t) =\mathcal{O}(V), \qquad \forall i,t.
\end{equation}

\item[(b)] \emph{\texttt{[Optimality]}}:The trajectory of Algorithm 1 is such that
\begin{equation}\label{Optimality}
\limsup\limits_{T\rightarrow\infty}\frac{1}{T}\sum_{t=1}^{T}\mathbb{E}\{{\rm BMSE}(t)\}\leq {\rm BMSE}^{\rm opt} + \frac{C+\Gamma}{V},
\end{equation}
where ${\rm BMSE}^{\rm opt}$ is the infimum Bayesian Mean-Square Error achievable by any policy that meets the required constraints in (\ref{Problem}).
\end{description}
\end{theorem}

\vspace{.1cm}
\begin{proof}
See Appendix A.
\end{proof}
\vspace{.1cm}

Theorem 1 illustrates that, increasing the penalty parameter $V$, the proposed method approaches the optimal solution of Problem (\ref{Problem}) [cf. (\ref{Optimality})], while guaranteeing stability of the battery levels and the energy causality constraint [cf. (\ref{batt1})]. The price to be paid to reach optimality is an increase of the required size of the batteries, which grows linearly with $V$ [cf. (\ref{batt2})]. In other words, Theorem 1 quantitatively expresses an existing tradeoff between decentralized estimation performance and size of the batteries at each node. Thus, once selected the battery size of each sensor, the parameter $V$ can be tuned to stabilize the battery level around a prefixed region of interest (e.g., around the maximum available size, or its 60\%-70\% for improved efficiency \cite{batt_univ}, as in most realistic settings), thus exploring the aforementioned tradeoff.

In the next paragraph, we illustrate some numerical results confirming our theoretical claims, and assessing the performance of the proposed method.

\begin{figure}[t]
	\centering
	\includegraphics[scale = .405]{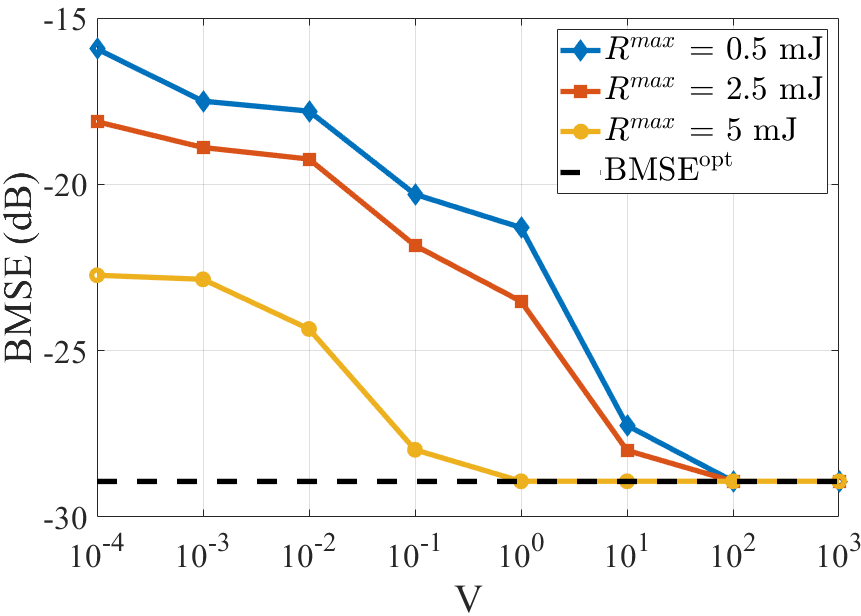}
	\caption{BMSE versus V, for different values of $R^{\max}$.}\label{Fig1}
\end{figure}

\subsection*{Numerical Results}

We consider a WSN with $N =50$ nodes uniformly distributed over a disk of radius 100 meters. The signal $\bx(t)$ is modeled as in \eqref{vector_obs} and it is defined over a graph whose adjacency matrix  considers Gaussian weights depending on the relative distance among the nodes \cite{shuman2013emerging}, using a scale parameter (i.e., the variance) $\alpha^2=0.25$. The resulting graph signal belongs to the subspace (i.e., matrix $\mU$ in (\ref{vector_obs})) spanned by the first six eigenvectors of the Laplacian matrix of the graph; the signal process \bs(t) in \eqref{vector_obs} is zero-mean with a randomly chosen covariance matrix $\mathbf{C}_{s}$ such that the worse BMSE in \eqref{BMSE_no_nodes}, corresponding to having an empty sampling set (i.e., no node is transmitting), is equal to -2 dB (loose prior). The observation noise in (\ref{observation}) is zero-mean, Gaussian, with a variance $\sigma_i^2=10^{-4}$ for all $i$. The radio channels $\{h_i(t)\}_{i,t}$ consider free-space propagation with a carrier frequency equal to $10$ MHz. We also added a multiplicative i.i.d. Rayleigh fading with unitary variance. The other parameters are: $A=1$, $G_d=10^{-3}$, $N_f=10$, ${\rm BER}_i=10^{-4}$ for all $i$. The slot duration is set to $T=1$ ms, and $e_i^{max}$ is chosen as in (\ref{power_QAM}) setting $b_i(t)=4$ bits (i.e., the maximum number of bits used for this simulation) for all $t$ and $i=1,\ldots,N$, and $h_i(t)$ selected as the fifth percentile of the radio channels' probability density function. The above setting has been chosen consistently with a realistic Bluetooth WSN \cite{blueWSN}. The harvested energies $\{R_i(t)\}_{i,t}$ are uniformly distributed between 0 and $R^{\max}$ for all $i,t$.

\begin{figure}[t]
\centering
	\includegraphics[scale = .395]{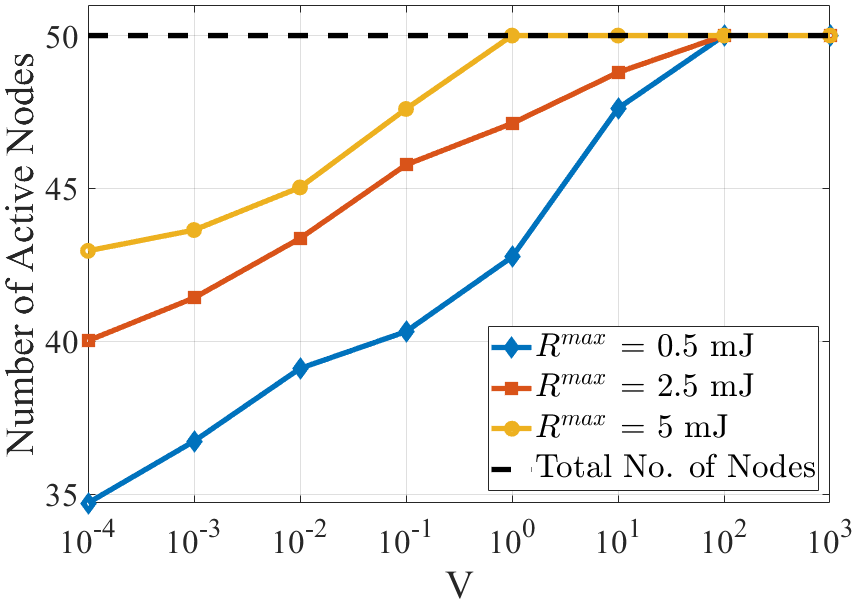}
	\caption{Active nodes versus V, for different values of $R^{\max}$.}\label{Fig2}
\end{figure}

\begin{figure}[t]
	\centering
	\includegraphics[scale = .4]{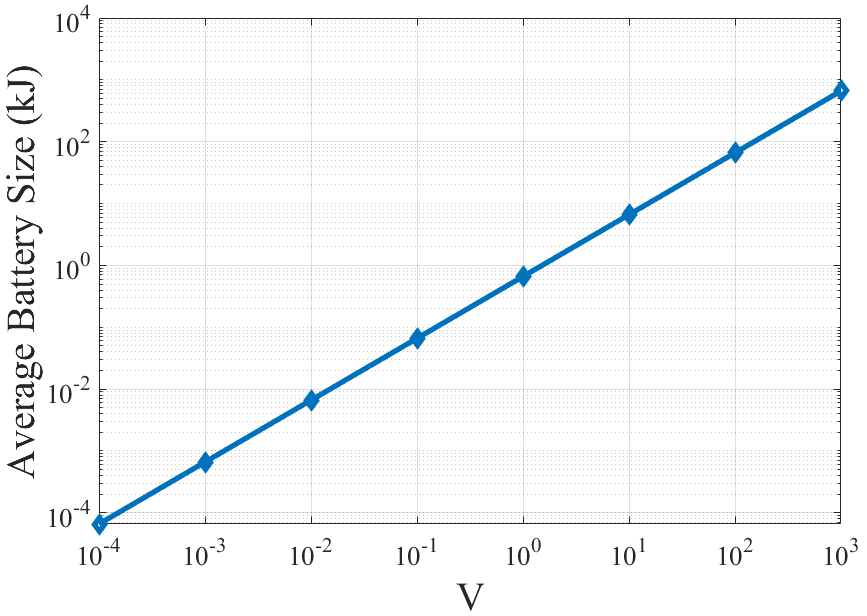}
	\caption{Battery level versus V.}\label{Fig3}
\end{figure}

To assess the performance of Algorithm 1, in Fig. \ref{Fig1} we illustrate the behavior of the average BMSE in (\ref{BMSE2}) versus $V$, averaged over time on 100 samples after convergence and over 50 independent simulations, considering different values of the maximum harvestable energy $R^{\max}$. We also report the minimum BMSE (i.e., $\mathrm{BMSE^{opt}}$) achievable by the system as a benchmark, which correspond to having all nodes active and transmitting using the maximum number of bits. As we can see from Fig. \ref{Fig1}, at large values of $V$, the proposed method approaches the optimal performance $\mathrm{BMSE^{opt}}$. As expected, increasing $R^{max}$ (i.e., the energy harvestable from the environment), the method gets to the optimal performance at lower values of $V$. Also, in Fig. \ref{Fig2}, we report the average number of active nodes (or, equivalently, the average cardinality of the sampling set) versus $V$, averaged over time and 50 independent simulations, for different values of  $R^{\max}$. From Fig. \ref{Fig2}, we can notice how the number of active nodes increases with $V$ and, as expected, the optimal ${\rm BMSE}^{\rm opt}$ in Fig. \ref{Fig1} corresponds to having all nodes active. Furthermore, in Fig. \ref{Fig3}, we illustrate the the average battery level of the network (on a logarithmic scale), which increases with $V$, having fixed $R^{\max} = 2.5 $ mJ. These results illustrate the tradeoff between estimation performance and size of the batteries, thus confirming the theoretical results in Theorem 1. In practical applications, this tradeoff imposes a limit on the achievable estimation performance that depends on the capacity of the batteries within each sensor, which is typically dictated by the economic cost to build the single device.

\begin{figure}[t]
	\centering
	\includegraphics[scale = .395]{nonstat_BMSE_png.png}
	\caption{BMSE versus time, with ON/OFF harvesting profile.}\label{Fig4}
\end{figure}

Finally, in Fig. \ref{Fig4}, we plot the temporal evolution of the BMSE in a non-stationary scenario, where we simulate an ON/OFF EH profile with a time window of 1 s. The results are averaged over 100 independent simulation, and consider $V = 10^{-4}$ and $R^{\max}$ = 5 mJ. As we can notice from Fig. \ref{Fig4}, the proposed procedure is able to react and adapt to the change in the EH statistics, achieving better or worse estimation performance depending on the current availability or absence of harvested energy, respectively.

\section{Dynamic Energy Minimization under Battery Stability and Estimation Accuracy Constraints}

In the previous section, we have focused our attention on a dynamic strategy that optimally selects radio parameters, the set of sampling sensors and the harvested energies in order to minimize the average BMSE while ensuring battery stability at each node. Often, specific applications impose constraints on the performance that the estimation procedure must satisfy. In this context, the problem can be formulated using a sparse sensing approach \cite{chepuri2016sparse}, i.e., as the minimization of the average energy spent by the WSN (and, as a byproduct, the number of transmitting nodes) to achieve the required value of estimation accuracy, while still guaranteeing strong stability of the batteries. The problem can be cast mathematically as:
\begin{equation}\label{Problem2}
\hspace{-0cm}\begin{array}{lll}
&\underset{\boldsymbol{e}(t),\boldsymbol{r}(t)} \min \quad  \displaystyle \lim_{t\rightarrow\infty}\,\frac{1}{t}\sum_{\tau=0}^{t-1}\,\sum_{i=1}^N\, {\mathbb{E}\left\{e_i(\tau)\right\}}\smallskip\\
& \quad\qquad \text{subject to} \\
&\quad \qquad \qquad \displaystyle \lim_{t\to \infty} \frac{1}{t} \sum_{\tau=0}^{t-1} \mathbb{E}\{B_i(\tau)\}<\infty \quad \forall i;\\
&\quad\qquad \qquad \displaystyle\lim_{t\to\infty}\,\frac{1}{t}\sum_{\tau=0}^{t-1}\, {\mathbb{E}\left\{\mathrm{BMSE}(\bee(\tau))\right\}}\leq \gamma; \smallskip\\
&\quad\qquad \qquad 0\leq e_i(t) \leq \min[e^{\max}_i,B_i(t)-e_{o,i}], \quad \forall i,t;\smallskip\\
&\quad\qquad \qquad\displaystyle 0\leq r_i(t) \leq R_i(t), \quad \forall i,t.
\end{array}
\end{equation}
The first constraint in (\ref{Problem2}) imposes that the battery evolution at each node is strongly stable; the second constraint imposes that the average BMSE is lower than a prescribed value $\gamma>0$; the third constraint puts bounds on the transmitted powers, i.e., the minimum among $B_i(t)-e_{o,i}$ and the maximum energy $e^{\max}_i$ that can be transmitted by the radio interface; finally, the last constraint in (\ref{Problem2}) sets the bounds on the harvestable energy at each time slot.

Similarly to the previous section, we use tools from stochastic optimization \cite{neely2010stochastic} to provide a dynamic strategy that solves problem (\ref{Problem2}).
To this aim, we first introduce the virtual queue $Z(t)$ associated with the BMSE inequality constraint in (\ref{Problem2}), with update equation:
\begin{align}\label{Z}
Z(t+1)=\max[Z(t)+\mu({\rm BMSE}(\bee(t))-\gamma),0],
\end{align}
where $\mu>0$ is a step-size used to control the convergence speed of the algorithm \footnote{The step-size does not alter the problem, and comes from the multiplication of both sides of the second constraint in (\ref{Problem2}) by a scalar $\mu>0$.}. Furthermore, to keep the energy storage stabilized around a maximum battery size, we use again the virtual queues \cite{neely2010dynamic,mao2016dynamic}:
\begin{align}
\widetilde{B}_i(t)=B_i(t)-\vartheta_i, \;\;\; i=1,\ldots,N, \label{Bi_tilde2}
\end{align}
where $B_i(t)$ evolves as in (\ref{battery}), and $\vartheta_i>0$ is a parameter to be selected. Then, we introduce the Lyapunov function:
\begin{equation}
L(\mathbf{\Psi}(t))=\frac{1}{2}Z(t)^2+\frac{1}{2}\sum_{i=1}^N \widetilde{B}_i(t)^2
\end{equation}
where $\mathbf{\Psi}(t)=\Big[Z(t),\{\widetilde{B}_i(t)\}_{i=1}^N\Big]$, and the one-slot conditional Lyapunov drift given by:
\begin{equation}\label{Lyapunov_drift}
\Delta(\mathbf{\Psi}(t))\triangleq\mathbb{E}\{L(\mathbf{\Psi}(t+1))-L(\mathbf{\Psi}(t))|\mathbf{\Psi}(t)\},
\end{equation}
where the expectation depends on the control policy, and is taken with respect to the random channels and energy packet arrivals. Minimizing (\ref{Lyapunov_drift}) would stabilize the virtual queues, but it can lead to a large energy expenditure. Since our approach aims at minimizing the energy spent by the network to perform the signal recovery task [cf. (\ref{Problem2})], we introduce the {\it drift-plus-penalty} function as \cite{neely2010stochastic}:
\begin{equation}\label{Lyapunov_drift_plus_penalty}
\Delta_p(\mathbf{\Psi}(t),\bee(t))=\Delta(\mathbf{\Psi}(t)) + V\,
\sum_{i=1}^N\, {\mathbb{E}\left\{e_i(t)\right\}}|\mathbf{\Psi}(t)\}
\end{equation}
where $V$ is a control parameter used to trade-off power consumption with queues length.
\begin{algorithm}[t]
\caption*{\textbf{Algorithm 2: Dynamic energy minimization under  battery stability and estimation accuracy constraints}}
\vspace{.1cm}
\textbf{Data:} Set $V>0$; $B_i(0)\geq 0 $, $Z(0)$ chosen positive and at random, $\theta_i >0$, for all $i=1,\dots,N$. Then, for each time $t\geq0$, observe the random events $\{h_i(t)\}_{i=1}^N$ and $\{R_i(t)\}_{i=1}^N$, and repeat:\smallskip
\begin{itemize}
\item \textbf{Fusion center:}  Compute the transmission energies $\{e_i(t)\}_{i=1}^N$ by solving: \vspace{-0.2cm}
	\begin{equation}\label{Dynamic_Problem}
	\hspace{-.65cm}\begin{array}{lll}
	&\displaystyle\underset{\boldsymbol{e}(t)} \min \;  \sum_{i=1}^N \left(V-\widetilde{B}_i(t)\right) e_i(t)+Z(t)\cdot{\rm BMSE}(\bee(t))\medskip\\
	&\quad\hbox{subject to} \;\; 0\leq e_i(t) \leq \min[e^{\max}_i,B_i(t)-e_{o,i}] \vspace{-0.1cm}
	\end{array}
	\end{equation}
Transmit $\{e_i(t)\}_{i=1}^N$ to the nodes.\smallskip

\item \textbf{Nodes:} Receive $\{e_i(t)\}_{i=1}^N$ from the fusion center. Nodes beloning to the sampling set $\mathcal{S}(t)$ sense data from the environment, and transmit them to the fusion center.
Update harvested energies $\{r_i(t)\}_{i=1}^N$ as (\ref{ri}). All nodes update batteries $\{B_i(t)\}_{i=1}^N$ as (\ref{battery}), and transmit them to the fusion center.\smallskip

\item \textbf{Fusion center:} Collect data from the sampling set $\mathcal{S}(t)$, and compute the LMMSE estimator in (\ref{lmmse}), where $m(y_i(t),b_i(t))$ and the $i$-th diagonal element of $\mC_w$ in (\ref{noise_covariance}) are set to 0 for all $i\notin \mathcal{S}(t)$. Update $Z(t)$ as in (\ref{Z}).
\end{itemize}
\end{algorithm}
Now, following arguments as in \cite{neely2010stochastic}, exploiting (\ref{Z}) and (\ref{Bi_tilde2}), the drift-plus-penalty function in (\ref{Lyapunov_drift_plus_penalty}) can be upper-bounded as:
\begin{align}\label{Lyapunov_drift_plus_penalty_bound}
&\Delta_p(\mathbf{\Psi}(t),\bee(t),\br(t))\,\leq\, C_2 + V\cdot\sum_{i=1}^N \mathbb{E}\{e_i(t)|\mathbf{\Psi}(t)\} \nonumber\\
&\qquad\quad+Z(t)\cdot\mathbb{E}\big\{{\rm BMSE}(\bee(t))-\gamma |\mathbf{\Psi}(t)\big\} \nonumber\\
&\qquad\quad+ \sum_{i=1}^N \widetilde{B}_i(t)\cdot\mathbb{E}\{  \br_i(t)-\bee_i(t)|\mathbf{\Psi}(t)\}
\end{align}
where $C_2$ is a positive constant. Thus, we can now use the stochastic approach where we greedily minimize instantaneous values of \eqref{Lyapunov_drift_plus_penalty_bound} at each $t$ \cite{neely2010stochastic}, thus obtaining the control policy described by Algorithm 2. The proposed dynamic algorithm determines the optimal transmission energies $\{e_i(t)\}_i$, the sampling set $\mathcal{S}(t)$, and the energies $\{r_i(t)\}_i$ to be harvested from the environment. In particular, minimizing \eqref{Lyapunov_drift_plus_penalty_bound} with respect to $r_i(t)$, with the constraint $0\leq r_i(t) \leq R_i(t)$,  each node $i$ collects the maximum harvestable energy $R_i(t)$ when $B_i(t)\leq\vartheta_i$; whereas, for $B_i(t)>\vartheta_i$, node $i$ does not harvest any energy [cf. (\ref{ri})]. Consequently, merging (\ref{battery}) with (\ref{ri}), we have:
\begin{equation}\label{Bound_batteries}
B_i(t)\leq\vartheta_i+R^{\max}-e_{o,i}, \qquad \hbox{for all $i,t$,}
\end{equation}
which guarantees the strong stability of the batteries required in (\ref{Problem2}). The control policy achieved by the proposed strategy, together with the overall decentralized estimation procedure in EH-WSNs, are listed in Algorithm 2.

The step in \eqref{Dynamic_Problem} requires the solution of a nonconvex optimization problem, which can be carried out using any descent approach initialized with a warm start, i.e., using the energy values $\{e_i(t-1)\}_{i=1}^N$ at time $t-1$. However, this approach has no guarantees to find the optimal solution of \eqref{Dynamic_Problem} at each time $t$, and in general has a large complexity that might not be affordable in real-time. Thus, proceeding as in Sec. III, we propose a simplified algorithm that has very low-complexity, while still guaranteing the BMSE constraint and offering energy performance that are close to those achieved by Algorithm 2. In particular, the approach is similar to the one exploited for Algorithm 1: the core idea is still finding a  suitable approximation  of  the  objective  function  in  (\ref{Dynamic_Problem}) in  order  to simplify its solution. Thus, we again substitute ${\rm BMSE}(\mathbf{e}(t))$ in (\ref{Dynamic_Problem}) with its linearization around $\mathbf{e}(t-1)$ (i.e., the solution available at time $t-1$), which is given by \eqref{Approx_penalty}. Using \eqref{Approx_penalty} in \eqref{Dynamic_Problem}, it is easy to see that the resulting problem is linear in $\{e_i(t)\}_{i=1}^N$, and its globally optimal solution is given in closed form by:
\begin{align}\label{ei2}
&e_i(t) = \min[e^{\max}_i,B_i(t)-e_{o,i}] \times \nonumber
\\&\;\;\qquad \mathbb{I}\left(\tilde{B}_i(t) \geq V+Z(t) \frac{\partial}{\partial e_i}{\rm BMSE}(\mathbf{e}(t-1))\right),
\end{align}
for all $t$ and $i=1,\ldots,N$. Clearly, the energy update in (\ref{ei2}) has much lower complexity than the one in (\ref{Dynamic_Problem}), which requires the solution of an optimization problem. In summary, the control policy achieved by the proposed low-complexity strategy, together with the overall decentralized estimation procedure in EH-WSNs, are listed in Algorithm 3. Deriving a full theoretical analysis for Algorithms 2 and 3 is a complicated task, and goes beyond the scope of this paper. The performance of the proposed strategies will be assessed through numerical experiments in the next section.

\begin{algorithm}[t]
\caption*{\textbf{Algorithm 3: Approximated Dynamic energy minimization under  battery stability and estimation accuracy constraints}}
\vspace{.1cm}
\textbf{Data:} Set $V>0$; $B_i(0)\geq 0$, $e_i(-1)$ chosen at random in $[0,e_i^{max}]$, $Z(0)$ chosen positive and at random, $\theta_i >0$, for all $i=1,\dots,N$. Then, for each time $t\geq0$, observe the random events $\{h_i(t)\}_{i=1}^N$ and $\{R_i(t)\}_{i=1}^N$, and repeat:\smallskip
\begin{itemize}
\item \textbf{Nodes:} Update harvested energies $\{r_i(t)\}_{i=1}^N$ as (\ref{ri}), and transmitted energies $\{e_i(t)\}_{i=1}^N$ as:
\begin{align}
&e_i(t) = \min[e^{\max}_i,B_i(t)-e_{o,i}] \times \nonumber
\\& \mathbb{I}\left(\tilde{B}_i(t) \geq V+Z(t)\cdot \frac{\partial}{\partial e_i}{\rm BMSE}(\mathbf{e}(t-1))\right) \,\forall i,t.  \nonumber
\end{align}
Nodes belonging to the sampling set $\mathcal{S}(t)$ sense data from the environment, and transmit them to the fusion center. All nodes update batteries $\{B_i(t)\}_{i=1}^N$ as (\ref{battery}).\smallskip
\item \textbf{Fusion center:} Collect data from the sampling set $\mathcal{S}(t)$, and compute the LMMSE estimator in (\ref{lmmse}), where $m(y_i(t),b_i(t))$ and the $i$-th diagonal element of $\mC_w$ in (\ref{noise_covariance}) are set to 0 for all $i\notin \mathcal{S}(t)$. Update $Z(t)$ as in (\ref{Z}). Evaluate and transmit $Z(t+1)\cdot\displaystyle\frac{\partial}{\partial e_i}{\rm BMSE}(\mathbf{e}(t))$ to node $i$, for all $i=1,\ldots,N$.
\end{itemize}
\end{algorithm}

 \begin{figure}[t]
	\centering
	\includegraphics[width=\columnwidth]{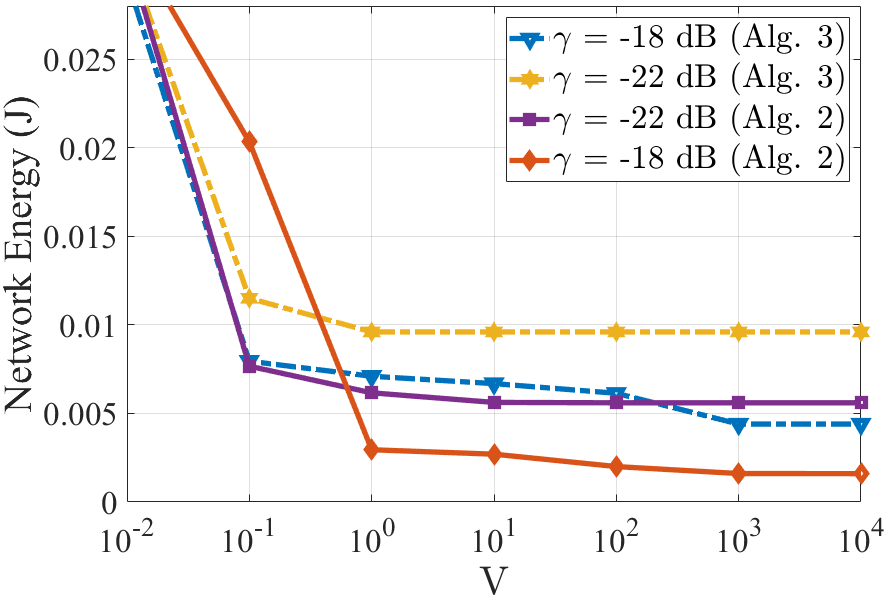}
	\caption{Transmitted energy versus V, for different values of the parameter $\gamma$.}\label{Fig5}
\end{figure}

\begin{figure}[t]
	\centering
	\includegraphics[width=\columnwidth]{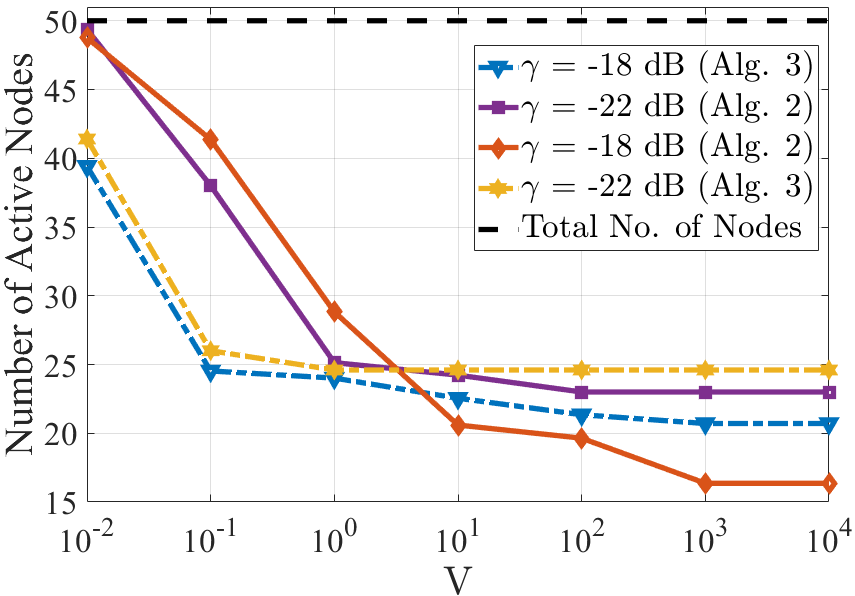}
	\caption{Number of active nodes versus V, for different values of the parameter $\gamma$.}\label{Fig6}
\end{figure}

\subsection*{Numerical Results}

\begin{figure}[t]
	\centering
	\includegraphics[width=\columnwidth]{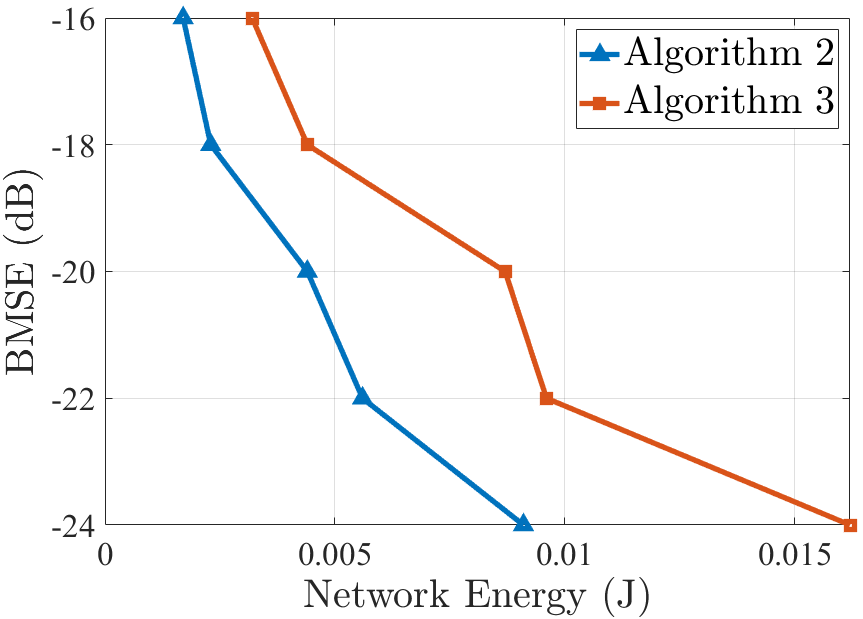}
	\caption{BMSE vs Network Energy, for Alg. 2 and Alg. 3.}\label{Fig11}
\end{figure}

In the sequel, we consider the same WSN and parameters configuration used in the previous case. In Fig. \ref{Fig5}, we illustrate the behavior of the sum of transmission energies over the network obtained by Algorithms 2 and 3, versus the control parameter $V$, for different values of the accuracy constraint $\gamma$. The results are averaged over time on 100 samples after convergence and over 50 independent simulations. From Fig. \ref{Fig5}, we can notice how the average transmission energy decreases at larger values of the trade off parameter $V$, until a floor value is reached. As expected, the floor value is higher when we require a stricter requirement on the estimation accuracy, i.e., at lower values of $\gamma$. Also, we can notice that Algorithm 2 shows better performance floors with respect to Algorithm 3. Interestingly, as a byproduct, increasing the parameter $V$, the solution of (\ref{Dynamic_Problem}) [and (\ref{ei2})] in terms of transmission energies tends to be always more and more sparse, i.e., many sensors do not transmit at all. This fact is quantified in Fig. \ref{Fig6}, where we illustrate the average behavior of the number of active nodes (or, equivalently, the average cardinality of the sampling set) with respect to $V$, for different values of $\gamma$ and different algorithms. The results are averaged over time on 100 samples after convergence and over 50 independent simulations. As we can see from Fig. \ref{Fig5}, the number of active nodes tends to decrease at large values of $V$, thanks to the sparsifying action of the control parameter. Also, Algorithm 2 enables a better sparsification of transmitting nodes thanks to its increased complexity. Finally, to compare the performance of Algorithms 2 and 3, in Fig. \ref{Fig11} we illustrate the trade-off between BMSE constraint and minimum network energy needed to achieve it. As we can see from Fig. \ref{Fig11},  Algorithm 2 produces a better BMSE-energy tradeoff with respect to Algorithm 3. However, the energy performance loss of Algorithm 3 is quite limited for all values of BMSE, while enabling a very large reduction of computational complexity.

\begin{figure}[t]
	\centering
	\includegraphics[width=\columnwidth]{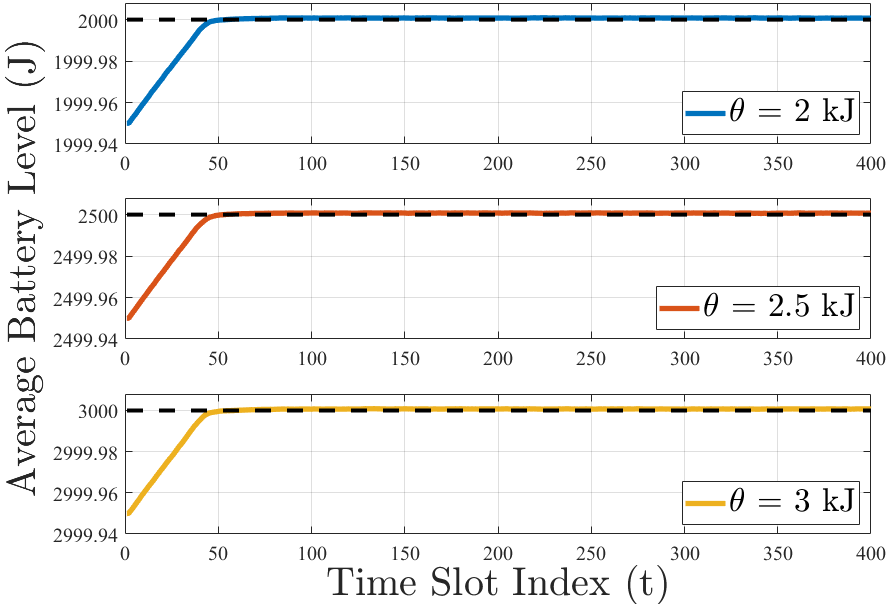}
	\caption{Average Battery Level versus time, for different values of the parameter $\vartheta_i$.}\label{Fig7}
\end{figure}

\begin{figure}[t]
	\centering
	\includegraphics[width=\columnwidth]{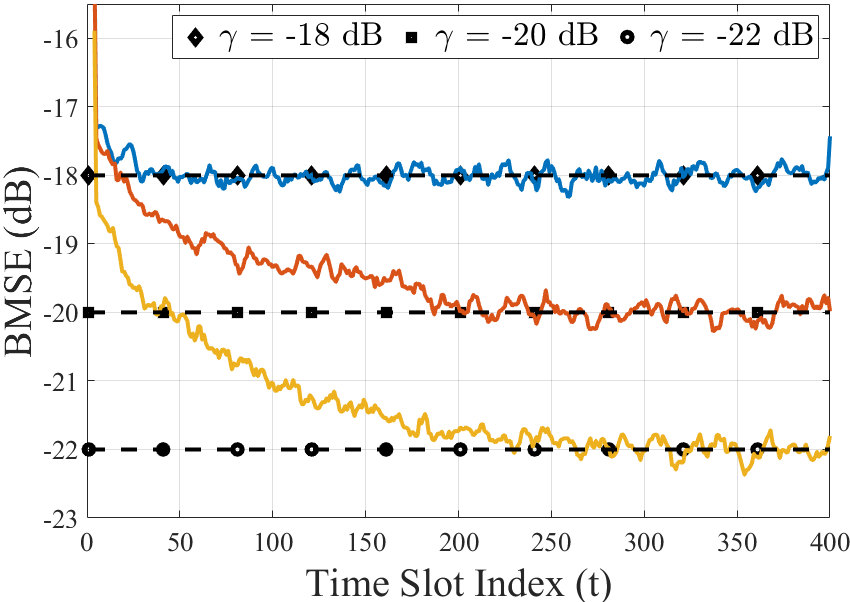}
	\caption{BMSE versus time, for different values of the parameter $\gamma$ using Algorithm 2.}\label{Fig8}
\end{figure}

\begin{figure}[t]
	\centering
	\includegraphics[width=\columnwidth]{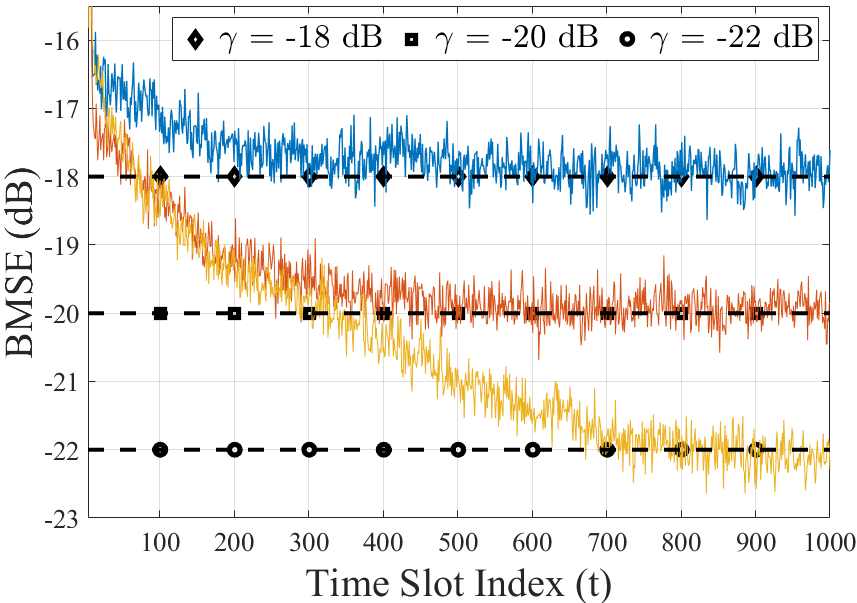}
	\caption{BMSE versus time, for different values of the parameter $\gamma$ using Algorithm 3.}\label{Fig9}
\end{figure}

\begin{figure}[t]
	\centering
	\includegraphics[width=\columnwidth, height = 6.2cm]{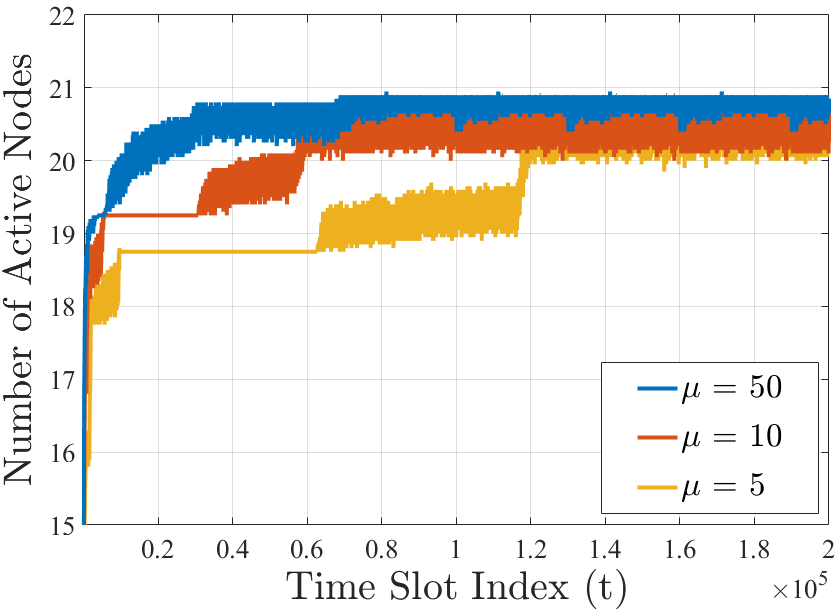}
	\caption{Number of Active Nodes versus time, for different values of the parameter $\mu$ using Algorithm 3.}\label{Fig10}
\end{figure}

The results in Figs. \ref{Fig5}, \ref{Fig6} and \ref{Fig11} are obtained while guaranteeing prescribed performance in terms of battery levels and estimation performance. In particular, in Fig. \ref{Fig7}, we illustrate the temporal behavior of the battery level, averaged over the network and over 50 independent simulations, considering $V=100$, $\gamma=-20$ dB and different values of $\vartheta_i=\vartheta$ for all $i=1,\ldots,N$. The step-size parameter $\mu$ was hand-tuned to obtain the best convergence rate in all the simulations. As we can notice From Fig. \ref{Fig7}, the battery levels quickly become stable around a value slightly greater than $\vartheta$, while satisfying the upper-bound in  (\ref{Bound_batteries}). In Figs. \ref{Fig8} and \ref{Fig9}, we show the temporal behavior of the BMSE for Algorithms 2 and 3, respectively, averaged over 100 independent simulations, for different values of $\gamma$. The simulation parameters are: $V=100$ and $\vartheta_i=$ 2.5 kJ for all $i$. As we can notice from Figs. \ref{Fig8} and \ref{Fig9}, both algorithms stabilize the BMSE around $\gamma$, thus guaranteeing the target performance of the estimation task. Finally, in Fig. \ref{Fig10} we show the temporal behavior of the number of active nodes using Algorithm 3, averaged over 50 independent simulations, for different values of $\mu$. The simulation parameters are: $V=100$, $\gamma = -18$ dB, $\vartheta_i=$ 2.5 kJ and $B_i(0) \approx \vartheta_i/2$ for all $i$. As expected, the convergence rate of the Algorithm depends on the virtual queue step-size $\mu$: larger values of $\mu$ force the optimization procedure to respect the constraint on the estimation performance as early as possible, thus enabling more nodes to transmit since the first time slots. Clearly, the convergence speed is also related to the battery level of the sensors: simulation results suggest not to overshoot the step-size $\mu$ if the initial battery level is low, because nodes would be forced to transmit slowing down the stabilization of the batteries.

\section{Conclusions}

In this paper, we have proposed dynamic strategies for optimal resource allocation in energy harvesting wireless sensor networks, aimed at performing decentralized (vector) signal estimation at a fusion center. The methods use a probabilistic digital quantization scheme to cope with the rate-constrained nature of the wireless channel.
The proposed strategies are based on stochastic Lyapunov optimization, and dynamically select radio parameters, sampling set, and harvested energy at each node, while guaranteeing accuracy of the recovery procedure, and stability of the batteries around a prescribed operating level. Interestingly, the strategy aimed at maximizing the BMSE introduces a trade-off between estimation accuracy and size of the battery levels at each node. The methods do not require any prior knowledge of the statistics of radio channel and/or harvested energy arrivals, and are capable to adapt in real-time to dynamic variations of the environment. Numerical results corroborate our theoretical findings, and assess the performance of the proposed strategies for decentralized signal estimation over energy harvesting WSNs. Future research directions include the application of the proposed techniques to enable general machine learning tasks (such as, e.g., prediction, anomaly detection,  data classification, etc.) at the edge of the wireless network.

\appendices
\section{Proof of Theorem 1}

\textbf{Point (a).} We proceed by analyzing separately all possible batteries states along the trajectory of Algorithm 1. Let us consider the battery's behavior at the $i$-th node.
\begin{itemize}
    \vspace{.1cm}
\item{ if $\theta_i \leq B_i(t) \leq \theta_i + R^{max}$ then:
\begin{align}
B_i(t+1) &= B_i(t) - e_i(t) + r_i(t)-e_{o,i} \nonumber\\
&= B_i(t) - e_i(t) + 0 -e_{o,i} \nonumber\\
& \leq B_i(t)-e_{o,i} \nonumber\\
& \leq \theta_i + R^{max}-e_{o,i}, \label{up_bbound_1}
\end{align}
where the second equality derives from (\ref{ri}).
}
\item{if $0 \leq B_i(t) \leq \theta_i$ then:
\begin{gather}
B_i(t+1) = B_i(t) - e_i(t) + r_i(t)-e_{o,i}   \nonumber
\\ \quad \leq B_i(t) + r_i(t)-e_{o,i} \nonumber
\\ \leq \theta_i + R^{max}-e_{o,i},\label{up_bbound_2}
\end{gather}
where the first inequality derives from $e_i(t)\geq 0,\forall i,t$; whereas the second inequality comes from $r_i(t)\leq R^{max}$, $\forall i,t$. Thus, from \eqref{up_bbound_1} and \eqref{up_bbound_2}, it follows:
\begin{align}
   & B_i(t) \leq \theta_i + R^{max}-e_{o,i} \quad \forall i,t. \label{upper_bound}
\end{align}
}
\item if $B_i(t) \geq 2e_i^{max}+2e_{o,i}$ then:
\begin{gather}
B_i(t+1) = B_i(t) - e_i(t) + r_i(t)-e_{o,i}  \geq e_i^{max}+e_{o,i} ,
\end{gather}
because $0\leq e_i(t) \leq e_i^{max}$, and $r_i(t) \geq 0$, $\forall i,t$.
\vspace{.1cm}
\item if $e_i^{max}+2e_{o,i} \leq B_i(t) \leq 2e_i^{max}+2e_{o,i}$, we impose $e_i(t) = 0$ (i.e., the node does not transmit) to be sure the constraint $ e_i(t) \leq B_i(t)-e_{o,i}$ holds for all $t$.
From (\ref{ei}), $e_i(t) = 0$ if:
\begin{align} &\tilde{B}_i(t)-V\cdot \frac{\partial}{\partial e_i}{\rm
BMSE}(\mathbf{e}(t-1)) \leq 0. \label{theta_cond}
\end{align}
Now, exploiting (\ref{Bi_tilde}) in (\ref{theta_cond}), we obtain:
\begin{align}
&B_i(t)-\theta_i-V\cdot \frac{\partial}{\partial e_i}{\rm BMSE}(\mathbf{e}(t-1))  \nonumber
\\ & \quad \qquad \,\leq\, 2e_i^{max}+2e_{o,i} -\theta_i + V \cdot G_i^{max} \leq 0
\end{align}
where $G_i^{max}$ is an upperbound of the (absolute value of the) $i$-th BMSE gradient component given by (\ref{Gimax}). The derivation of $G_i^{max}$ is reported in Appendix 2. Thus, choosing:
\begin{gather}\label{theta_bound}
\theta_i \geq VG_i^{max} + 2e_i^{max}+2e_{o,i}
\end{gather}
ensures that $e_i(t) = 0$ and
 \begin{align}\label{lower_bound}
&B_i(t+1) = B_i(t) - e_i(t) + r_i(t)-e_{o,i}   \nonumber
\\& \qquad\qquad = B_i(t) - 0 + r_i(t)-e_{o,i}  \nonumber
\\& \qquad\qquad\geq e_i^{max}+e_{o,i} .
 \end{align}
In conclusion, if $B_i(0) \geq e_i^{max}+2e_{o,i}$ for $i = 1,...,N$, from (\ref{lower_bound}) and (\ref{upper_bound}), we obtain:
\begin{gather}\label{battery_bounds}
e_i^{max}+e_{o,i}  \leq B_i(t) \leq \theta_i+R^{max}-e_{o,i}, \quad\forall i,t,
\end{gather}
which proves (\ref{batt1}). Finally, choosing $\theta_i = VG^{max} + 2e_i^{max}+2e_{o,i}$ [cf. (\ref{theta_bound})], it follows from (\ref{battery_bounds}) that:
\begin{gather} \label{b_ord_v}
B_i(t) =\mathcal{O}(V), \qquad \forall i,t,
\end{gather}
which proves also (\ref{batt2}).
\end{itemize}

\textbf{Point (b).} The proof follows from the fact that the control policy given by Algorithm 1 is the solution of \eqref{main_prob}, which is a $\Gamma$-additive approximation of the drift-plus-penalty algorithm in \eqref{nonconv_prob}. This holds true because, for any given batteries state  $\mathbf{\tilde{B}}(t)$ at slot $t$, the objective function in \eqref{nonconv_prob} is bounded from above inside the finite size of the feasible set $\mathcal{Z}(t)$ in (\ref{set_Z}), for all $t$. Thus, the  conditional expected value of the objective function in \eqref{nonconv_prob}, evaluated in the solution of \eqref{main_prob}, is within a constant $\Gamma$ from the global optimum of problem \eqref{nonconv_prob}. Furthermore, from (\ref{battery_bounds}), the trajectory of (\ref{main_prob}) is always feasible for Problem (\ref{Problem}) [and, thus, (\ref{nonconv_prob})]. Then, the main claim comes as a direct consequence of \cite[Th. 4.8]{neely2010stochastic}.

\section{Derivation of $G_i^{max}$ in (\ref{Gimax})}
Using basic rules of matrix differentiation to (\ref{BMSE}) \cite{petersen2008matrix}, the ciclic property of the trace, and letting
 \begin{gather}\label{matrixL}
 \mathbf{L}(\bee(t)) =  \mathbf{C}_{ss}^{-1} + \sum_{i = 1}^N \frac{e_i^2(t)}{e_i^2(t)\sigma_i^2+A^2c_i^2(t)}
 \bu_i\bu_i^T,
 \end{gather}
the $i$-th component of the gradient of the BMSE is given by:
\begin{gather}\label{partial1}
 \frac{\partial }{\partial e_i}\mathrm{BMSE}(\bee(t)) = -{\rm Tr}\left\{\mathbf{L}^{-2}(\bee(t)) \frac{\partial \mathbf{L}(e_i(t))}{\partial e_i}\right\},
 \end{gather}
 where
 \begin{gather}\label{partial2}
 \frac{\partial \mathbf{L}(e_i(t))}{\partial e_i} = \underbrace{\frac{2e_i(t)A^2c_i^2(t)}{(e_i^2(t)\sigma_i^2 + A^2c_i^2(t))^2}}_{h(e_i(t),c_i^2(t))}\bu_i \bu_i^T.
 \end{gather}
 The algorithm trajectory in (\ref{ei}) determines that $e_i(t)$ can assume only the values 0 or $e_i^{max}$ [cf. (\ref{ei})], so $h(e_i(t),c_i^2(t))$ can obviously be upper bounded as:

  \begin{gather}\label{partial3}
\frac{2e_i(t)A^2c_i^2(t)}{(e_i^2(t)\sigma_i^2 + A^2c_i^2(t))^2} \leq \underbrace{\frac{2e_i^{max}A^2c_i^2(t)}{((e_i^{max})^2\sigma_i^2 + A^2c_i^2(t))^2}}_{g(c_i^2(t)) \overset{\Delta}{=} h(c_i^2(t),e_i^{max})}.
 \end{gather}
It is then easy to see that $g(c_i^2(t))$, clearly defined on positive real numbers,  has a maximum in the argument:
\begin{align} \label{max_g}
    c_i^2(t) = \frac{(e_i^{max})^2\sigma_i^2}{A^2}.
\end{align}
Now, the $i$-th gradient component is different from 0 only if $e_i(t)=e_i^{max}$. Furthermore, from (\ref{matrixL})-(\ref{partial2}), we have that the worst-case (in terms of gradient magnitude) is achieved when node $i$ transmits, i.e., $e_i(t)=e_i^{max}$, while all other nodes are idle, i.e., $e_j(t)=0$, for all $j\neq i$.
Thus, setting $c_i(t)^2= 0$ in \eqref{matrixL} and plugging \eqref{max_g} in the RHS of \eqref{partial3}, an upper-bound of the absolute value of (\ref{partial1}) is given by:
\begin{align}
    &G_i^{max}=  \frac{1}{2e_i^{max}\sigma_i^2} {\rm Tr}\left\{ \left(\mathbf{C}_{s}^{-1} + \sigma_i^{-2}\bu_i\bu_i^T\right)^{-2}\hspace{-.1cm}\bu_i\bu_i^T  \right\}, \nonumber
\end{align}
for all $i=1,\ldots,N$.

\balance
\bibliographystyle{IEEEbib}
\bibliography{refs}
\end{document}